\documentclass[12pt]{article} 
\usepackage{amsmath,amssymb} 
\usepackage{cite}
\usepackage[unicode,linktocpage=true,plainpages=false,pdfpagelabels=false]{hyperref}

\begin{document}

\title{Description of gravity in the model with independent nonsymmetric connection}

\author{N.~V.~Kharuk\thanks{natakharuk@mail.ru}, S.~A.~Paston\thanks{ s.paston@spbu.ru}, A.~A.~Sheykin \thanks{a.sheykin@spbu.ru}\\
{\it Saint Petersburg State University, Saint Petersburg, Russia}
}
\date{\vskip 15mm}
\maketitle

\abstract{A generalization of General Relativity is studied. 
The standard Einstein-Hilbert action  is considered in the Palatini formalism, where the connection and the metric are independent variables, and the connection is not symmetric. As a result of variation with respect to the metric Einstein equations are obtained. A variation with respect to the connection leads to an arbitrariness in the determination of connection, i.e. the presence of gauge invariance. Then a matter in a form of point particle which interacts with field of connection is introduced. Also the action 
is complemented by a kinetic term for field of the connection to avoid incompatible equation of motion. 
Thus after the variation procedures we obtain the Einstein equations, the geodesic equation 
and the Maxwell's equations for electromagnetism, where some components of the connection  play the role of the electromagnetic potential. Thereby the electromagnetic potential is obtained from the geometry of space-time.
}

\newpage
\maketitle
\section{Introduction}
\label{intro}
Nowadays the General Relativity is a modern theory of gravity. It is described
by curved 4-dimential  pseudo-Riemannian space-time. Usually in the framework of this theory the main objects are the symmetric 
metric $g_{\mu\nu}$ and the connection $\Gamma^\xi_{\mu \nu}$, which also symmetric and depends on the metric. The equivalent approach
is the so-called Palatini formalism in which the metric and the connection are considered as independent variables, but both are still symmetric. 

In this work we consider a generalization 
of this theory. The connection is considered as nonsymmetric and independent of the metric object. The metric is symmetric as usual. 
In these assumptions connection $ \Gamma^\alpha_{\rho \nu} $ is not equal to the Christoffel symbol. It  leads to the fact that such 
theory is more extensive. For example, the contraction $R^\alpha{}_{\alpha\mu\nu} $ is not equal to zero and the Ricci curvature tensor $R_{\beta\nu}\equiv R^\alpha{}_{\beta\alpha\nu}$ is not symmetric anymore.

 Such approach with the Einstein-Hilbert action for the first time was proposed only in 1978 in the paper \cite{Krechet}, which was published only in Russian and remained practically unknown to the scientific community.
 The authors noticed that this theory can be interpreted as unified theory of gravity and electromagnetism, however they did not introduce an action for the matter. Instead the authors  have to used some \textit{ad hoc} physical assumptions to obtain correct equations of motions of test particles. In the same year a very similar theory was considered in the paper \cite{Hehl}. However the purpose of the authors was different: they did not try to connect their theory with electromagnetism. Nowadays the studies of Palatini formalism with nonsymmetric connection have been continuing, see. e.g. \cite{bernal2017}. 
 
In the present paper we generalize the approach proposed in \cite{Krechet} by including a matter in the form of point particles into the action. We obtain a correct form of the action term for the interaction of such a matter with the connection and discover that it is necessary to identify the electromagnetic potential with the trace of the connection rather with the trace of the torsion (as it was done in \cite{Krechet}).

\section{Theory without matter}
To begin with, let us consider the Einstein-Hilbert action without matter in a standard form:

\begin{equation}\label{eq1}
S_1=-\frac{1}{2\varkappa} \int d^4x \sqrt{-g} g^{\mu\nu} R_{\mu\nu}(\Gamma),
\end{equation}
where 
 $R_{\beta \nu} = R^\alpha{}_{   \beta\alpha\nu} = \partial_\alpha \Gamma^\alpha_{\nu\beta}- \partial_\nu \Gamma^\alpha_{\alpha\beta} +
 \Gamma^\alpha_{\alpha\xi}\Gamma^\xi_{\nu\beta}- \Gamma^\alpha_{\nu\xi}\Gamma^\xi_{\alpha\beta}$ is the Ricci tensor which depends only on connection.

To obtain equations of motion it is necessary to vary the action. It is easy task to vary with respect to the metric $ g_{\mu\nu}$ because now action does not depend on the derivative of metric. Thereby one can get 
\begin{equation}\label{eq2}
 R^{\nu \mu}+R^{\mu \nu} -R g^{\mu \nu}=0.
\end{equation}
These equations differ from Einstein equations without matter because here $R^{\mu\nu}$ is not a symmetric tensor, so it contains an additional symmetrization. 
 
 Variation with respect to the connection gives more interesting result:
\begin{equation}\label{eq3}
 D_\rho g^{\sigma \nu}=\frac{1}{3}g^{\sigma \nu}S^\alpha_{\rho \alpha}+\frac{1}{3}g^{\sigma \xi}S^\alpha_{\xi \alpha}\delta^\nu_\rho +g^{\xi \sigma}S^\nu_{\rho \xi},
\end{equation}
where $S^\rho_{\mu \nu} = \Gamma^\rho_{\mu \nu}- \Gamma^\rho_{\nu \mu}$ is a torsion.
This equation is solved by
\begin{equation}\label{eq4}
 \Gamma^\rho_{\mu \nu}=\bar{\Gamma}^\rho_{\mu \nu}+\frac{1}{4}A_\mu \delta^\rho_\nu,
\end{equation}
where $\bar{\Gamma}^\rho_{\mu \nu} = \frac{1}{2} g^{\alpha \sigma}(\partial_\rho g_{\sigma \nu}+\partial_\nu g_{\rho \sigma}-
\partial_\sigma g_{\nu \rho})$ is the Christoffel symbols and
$A_\mu$ is an arbitrary vector. This formula can be rewritten  in a form:
\begin{equation}\label{eq5}
 \Gamma^\rho_{\mu \nu}-\frac{1}{4}\Gamma^\alpha_{\mu\alpha}\delta^\rho_\nu=\bar{\Gamma}^\rho_{\mu \nu}-\frac{1}{4}\bar{\Gamma}^\alpha_{\mu \alpha}\delta^\rho_\nu,
\end{equation}
by substitution trace of relation (\ref{eq4}). Now it is clear that the trace of the connection has not any restrictions. 
In this sense, this theory has additional gauge symmetry. 

Taking  (\ref{eq4}) into account  the explicit form of the curvature tensor is 
\begin{equation}\label{eq6}
R^\mu{}_{\nu\lambda\rho}=\bar{R}^\mu{}_{\nu\lambda\rho}+\frac{1}{4}(\partial_\lambda A_\rho - \partial_\rho A_\lambda)\delta^\mu_\nu,
\end{equation}
where $\bar{R}^\mu{}_{\nu\lambda\rho}$ is the usual Riemann curvature tensor which depends on the Christoffel symbols $\bar{\Gamma}^\rho_{\mu \nu}$ and thus can be expressed through metric. Taking trace of (\ref{eq6}) one can find that
\begin{equation}\label{eq7}
R^\mu{}_{\mu\lambda\rho}=\partial_\lambda A_\rho - \partial_\rho A_\lambda
\end{equation}
and
\begin{equation}\label{eq8}
R^\mu{}_{\lambda\mu\rho}\equiv R_{\lambda\rho}=\bar{R}_{\lambda\rho}+\frac{1}{4}(\partial_\lambda A_\rho - \partial_\rho A_\lambda),
\end{equation}
where $\bar{R}_{\lambda\rho}$ is a Riemannian and therefore symmetric Ricci tensor.
The scalar curvature:
\begin{equation}\label{eq9}
R\equiv R_{\lambda\rho}g^{\lambda\rho}=\bar{R}_{\lambda\rho}g^{\lambda\rho}
\end{equation}
does not change. Thereby the equation (\ref{eq2}) can be rewritten as:
\begin{equation}\label{eq10}
 \bar{R}^{\nu \mu} -\frac{1}{2}R g^{\mu \nu}=0.
\end{equation}
So it is exactly the vacuum Einstein's equations. 

\section{Addition of matter}

The next step is an addition of a matter in the theory. We will consider it in the form of a set of relativistic point particles. For the sake of simplicity we will write the formulas for a single particle. This particle with a world line ${x}^\mu(\tau)$ in the gravitational field with the metric $g_{\mu\nu}$ is described by standard action:
 \begin{equation}\label{eq11}
  S_2= -m\int d\tau \sqrt{\dot{x}^\mu(\tau) \dot{x}^\nu(\tau) g_{\mu \nu}(x(\tau))},
 \end{equation}
where $m$ is a mass of particle. 

In order to get a more general theory the interaction of classical particles with the  connection is introduced. One of the most simplest way is
 \begin{equation}\label{eq12}
  S_3=-q \int d\tau \dot{x}^\mu(\tau) \Gamma^\nu_{\mu \nu}(x(\tau)),
 \end{equation}
where $q$ is just a parameter\footnote{Another simple combination is the contraction of $\dot{x}^\mu$ with the trace of torsion $S^\nu_{\mu \nu}$ instead of the trace of connection $\Gamma^\nu_{\mu \nu}$. We will discuss it in the section \ref{disc}.}. 

Thus  the theory consists of three terms $S_1+S_2+S_3$. But it turns out that such theory is self-inconsistent. 
Indeed, after a variation with respect to the connection the following equations of motion arise:
\begin{equation}\label{eq13}
\dot{x}^\mu(\tau)=0
\end{equation}
which are incompatible with the normalization of four-velocity. This result is connected with the presence of gauge invariance in $S_1$ (see after (\ref{eq5})). To avoid this problem one can add one more term to the total action. It is the kinetic term which can be written in different ways, but again the most simple case is chosen:
\begin{equation}\label{eq14}
 S_4=-\frac{1}{16\pi} \int d^4 x \sqrt{-g}R^\mu{}_{\mu \alpha \beta}R^\mu{}_{\mu \delta \gamma} g^{\alpha \delta}g^{\beta \gamma},
\end{equation}
where the constant $\frac{1}{16\pi}$  is chosen for convenience. It is worth noting that the same term was proposed in \cite{Krechet}, but without above motivation since the action of matter was not considered in their work.

As a result the total action consists of four terms:
\begin{multline}\label{eq15}
 S=S_1+S_2+S_3+S_4=-\frac{1}{2\varkappa} \int d^4x \sqrt{-g}R - m\int d\tau \sqrt{\dot{x}^\mu(\tau) \dot{x}^\nu(\tau) g_{\mu \nu}}(x(\tau))-\\
 -q \int d\tau \dot{x}^\mu(\tau) \Gamma^\nu_{\mu \nu}(x(\tau))-\frac{1}{16\pi} \int d^4 x \sqrt{-g}R^\mu{}_{\mu \alpha \beta}R^\mu{}_{\mu \delta \gamma} g^{\alpha \delta}g^{\beta \gamma}.
\end{multline}
This is the final view of the action. It depends on three independent variables: the metric $g_{\mu\nu}$, the connection $\Gamma^\alpha_{\beta\gamma}$ and 
the coordinate of the particle $x^\mu(\tau)$.

Let us obtain the complete set of field equations by varying with respect to these variables. Firstly, let us consider the variation with respect to the metric. Only terms $S_1,$  ~ $ S_2$ and $S_4$ depend on $g_{\mu\nu}.$ 
The contribution of $S_1$ is already found in \eqref{eq10}. The variation of $S_2$ and $S_4$ can be easily calculated. Finally equations 
of motion corresponding to the variation of the metric are:
\begin{equation}\label{eq16}
 \bar{R}^{\mu \nu}-\frac{1}{2}R g^{\mu \nu} = \varkappa (T_1^{\mu \nu} + T_2^{\mu\nu}),
\end{equation}
where 
\begin{equation}\label{eq17}
  T_1^{\mu \nu}=\rho_m u^\mu u^\nu 
  \end{equation}
is a stress-energy tensor of a relativistic particle,
\begin{equation}\label{eq18}
  T_2^{\mu \nu}=- \frac{1}{4\pi} (R^\xi{}_\xi{}^{\mu \alpha} R^\beta{}_\beta{}^\nu{}_\alpha-\frac{1}{4}g^{\mu\nu}R^\xi{}_\xi{}^{\alpha\beta}R^\varphi{}_{\varphi\alpha\beta}),
 \end{equation} 
$u^\mu=\dot{x}^\mu \frac{1}{\sqrt{\dot{x}^\alpha\dot{x}^\beta g_{\alpha\beta}}}$ is a four-velocity and $\rho_m =m \int ds \delta(x-x(s))\frac{1}{\sqrt{-g(x(s))}}$ is a mass density.
 The expression for $T_2^{\mu \nu} $ can be rewritten in a more recognizable form using notation $R^\varphi{}_{\varphi\alpha\beta} \equiv F_{\alpha\beta}$:
 \begin{equation}\label{eq19}
  T_2^{\mu \nu}=- \frac{1}{4\pi} (F^{\mu \alpha} F^\nu{}_\alpha-\frac{1}{4}g^{\mu\nu}F^{\alpha\beta}F_{\alpha\beta}).
 \end{equation}
In this form $T_2$ reproduces the stress-energy tensor of electromagnetic field with electromagnetic tensor $F_{\alpha\beta}$.

Next the connection $\Gamma^\alpha_{\mu\nu} $ is varied. The term $S_2$ does not depend on the connection. The result 
for $S_1$ is already found in (\ref{eq3}). After calculations for $S_3$ and $S_4$ the following equation is obtained:
\begin{gather}\label{eq21}
  \Gamma^\rho_{\mu \nu}=\bar{\Gamma}^\rho_{\mu \nu}+\frac{1}{4}A_\mu \delta^\rho_\nu,\\\label{eq21'}
  \bar{D}_\mu F^{\mu\nu}=4\pi j^\nu,
 \end{gather}
where $\bar{D}_\mu$ is a covariant derivative with the Riemannian connection and
$j^\nu=q \int ds u^\nu \delta(x-x(s))\dfrac{1}{\sqrt{-g(x(s))}}$ is a four-current of relativistic particles if $q$ is considered as an electric charge. The equation \eqref{eq21'} is nothing but the Maxwell's equation. According to \eqref{eq21} and its consequence \eqref{eq7}  the role of electromagnetic potential corresponding to $F^{\mu\nu}$ is played by the quantity $A_\rho$.

Finally, the variation with respect to the particle coordinate is calculated. It is necessary to vary only $S_2$ and $S_3$. The result is already known since $S_2$ has a standard view and $S_3$ can be treated as interaction term for a particle with electromagnetic potential $\Gamma^\alpha_{\beta\alpha}$.
So the corresponding equation of motion is
\begin{equation}\label{eq20}
m u^\mu \bar{D}_\mu u^\alpha= -qu_\xi F^{\xi\alpha}.
\end{equation}
Thus this equation reproduces the equations of motion of relativistic particle in the gravitational field with metric $g_{\mu\nu}$ and the electromagnetic field again defined by the quantity $A_\rho$ in (\ref{eq7}) as a potential.

As a result we conclude that the system of field equations \eqref{eq16}, \eqref{eq21}, \eqref{eq21'}, \eqref{eq20} corresponding to action \eqref{eq15} reproduce Einstein-Maxwell equations. 
\section{Discussion}\label{disc}
Thereby only geometric objects were introduced such as the metric $g_{\mu\nu}$ and the connection $\Gamma^\alpha_{\mu\nu}$. 
The standard Einstein-Hilbert action was complemented by additional terms, which corresponding to a
relativistic point-like particle (\ref{eq11}), an interaction (\ref{eq12}) and a kinetic term (\ref{eq14}).
The additional degrees of freedom of the connection $A_\mu$ were treated as an electromagnetic potential.
As a result, the electrodynamics in gravitational field is constructed.
Despite the simplicity of the above theory, is had not been discovered at the times of the most intense search of unified field theory (for the detailed historical survey see \cite{statja47} and the references therein).

As we said above, for the first time a similar approach was proposed in 1978 \cite{Krechet}. However, they treated the trace of the \textit{torsion} as the electromagnetic potential. If we restrict ourselves to consideration of the matter-free action consisting only of  $S_1$ and $S_4$, the difference between identification of electromagnetic potential with the trace of the torsion and with the trace of connection (as in our approach) turns out to be negligible. However, the addition of matter changes the picture drastically. While the matter can be coupled with the trace of connection without any troubles, coupling with the trace of torsion
 \begin{equation}\label{eq22'}
  S'_3=-q \int d\tau \dot{x}^\mu(\tau) S^\nu_{\mu \nu}(x(\tau))
 \end{equation}
leads to the appearance of matter in the expression for the connection. Instead of  \eqref{eq21} we have
\begin{equation}\label{eq22}
 \Gamma^\rho_{\mu \nu}=\bar{\Gamma}^\rho_{\mu \nu}+\frac{1}{3} A_\mu \delta^\rho_\nu+ \frac{\varkappa}{4}\left({3} j^\rho g_{\mu\nu}  -  j_\mu \delta^\rho_\nu - j_\nu \delta^\rho_\mu\right).
\end{equation}
Note that the authors of \cite{Krechet} did not consider the variational principle for the matter. Instead they used some additional physical assumptions in order to obtain the equations of motion of test particles which are identical to \eqref{eq20}. 
It is worth noting that the paper \cite{Krechet} was published in the obscure Soviet journal which is almost completely unavailable to the scientific community, so the result remained unknown. The theory proposed in \cite{Krechet} was later rediscovered many times, e.g. in \cite{Hehl}, where the authors did not consider the interpretation of the $A_\mu$ as an electromagnetic potential, and in \cite{tucker1995}, where it served as the base for the possible extension of GR.

In the present work the role of matter is played by point particles. The generalization on the case of ideal fluid (i. e. continuous medium consisting on classical particles) is quite simple. The corresponding interaction term was proposed in \cite{statja47} for a variant of description of ideal fluid which was studied in \cite{statja48}.  
However, the interpretation of the $A_\mu$ as an electromagnetic potential is possible only for the purely classical description of matter. The generalization such theory for the more physical interesting case, where a matter is considered as a field, is a one way of the further development. For this purpose one could consider the approach of frame bundles. Such attempts, for example, was made by Horie \cite{horie2}. 

{\bf Acknowledgements.}
The work was supported by SPbGU grant N~11.38.223.2015.

\providecommand{\eprint}[1]{\href{http://arxiv.org/abs/#1}{\texttt{#1}}}










\end{document}